**Interlayer Exciton–Phonon Bound State in Bi$_2$Se$_3$/monolayer WS$_2$ van der Waals Heterostructures**

Zachariah Hennighausen[1,*], Jisoo Moon[1], Kathleen M. McCreary[2], Connie H. Li,[2] Olaf M.J. van 't Erve[2], and Berend T. Jonker[2,*]

[1] NRC Postdoc at the Materials Science and Technology Division, Naval Research Laboratory, Washington, D.C. 20375, USA

[2] Materials Science and Technology Division, Naval Research Laboratory, Washington, D.C. 20375, USA

**Abstract**

**The ability to assemble layers of two-dimensional (2D) materials to form permutations of van der Waals heterostructures provides significant opportunities in materials design and synthesis. Interlayer interactions provide a path to new properties and functionality, and understanding such interactions is essential to that end. Here we report formation of interlayer exciton-phonon bound states in Bi$_2$Se$_3$/WS$_2$ heterostructures, where the Bi$_2$Se$_3$ A$_1^{(3)}$ surface phonon, a mode particularly susceptible to electron-phonon coupling, is imprinted onto the excitonic emission of the WS$_2$. The exciton-phonon bound state (or exciton-phonon quasiparticle) presents itself as evenly separated peaks superposed on the WS$_2$ excitonic photoluminescence spectrum, whose periodic spacing corresponds to the A$_1^{(3)}$ surface phonon energy. Low-temperature polarized Raman spectroscopy of Bi$_2$Se$_3$ reveals intense surface phonons and local symmetry breaking that allows the A$_1^{(3)}$ surface phonon to manifest in otherwise forbidden scattering geometries. Our work advances knowledge of the complex interlayer van der Waals interactions, and facilitates technologies that combine the distinctive transport and optical properties from separate materials into one device for possible spintronics, valleytronics, and quantum computing applications.**

* Authors for correspondence, E-mail:  hennigha@mit.edu; berry.jonker@nrl.navy.mil;

**Introduction**

Van der Waals heterostructures are formed by stacking monolayers of 2D materials in any sequence of one's choosing, enabled by the lack of bonds between layer planes.[1] Such stacking often results in new properties, tuned by either material selection or twist angle.[2] Understanding the interaction between the layers in such heterostructures is essential to discovering new properties, engineering functionality, and advancing their application into technologies. Previous work showed the interlayer interaction between stacked two dimensional (2D) materials can facilitate a host of new properties, including long-lived interlayer excitons,[3] magnetic phase switching,[4] forbidden Raman modes,[5] superconductivity,[6] orbital ferromagnetism,[7] and emergent ferromagnetism.[8]

In this work, we grew few-layer $Bi_2Se_3$ on monolayer $WS_2$, and observe interlayer exciton-phonon coupling and formation of an exciton-phonon bound state between localized excitons in monolayer $WS_2$ and the $Bi_2Se_3$ $A_1^{(3)}$ surface phonon, a mode particularly susceptible to electron-phonon coupling.[9,10] The bound state is manifested as a series of evenly spaced peaks superposed on the $WS_2$ excitonic photoluminescence (PL) spectrum, whose periodic spacing corresponds to the $A_1^{(3)}$ surface phonon energy. Oscillating features that match a phonon energy and correspond to a luminescence are indicative of electron-phonon or exciton-phonon bound states.[11–15] In addition, polarized Raman spectroscopy of the $Bi_2Se_3$ reveals multiple pronounced surface phonon modes and crystalline symmetry breaking. Notably, the presence of surface phonons in a forbidden scattering geometry suggests local symmetry breaking at the surface,[9] consistent with a strong $WS_2$-$Bi_2Se_3$ interlayer coupling. Previous work found significant interlayer hybridization in $Bi_2Se_3$/$WS_2$ heterostructures, facilitating electron transfer and modifying the bonding,[16–18] conditions which encourage the formation of interlayer quasiparticles.[19] Understanding the interlayer interaction is central to elucidating how their combined properties evolve, enabling the discovery of advanced capabilities for spintronics,[18,20] valleytronics,[21] and quantum computing[21,22] applications.

Several publications have reported exciton-phonon or electron-phonon coupling across the interlayer region, where a free exciton in a monolayer transition metal dichalcogenide (TMD) facilitates the emergence of otherwise forbidden Raman modes in the adjacent material.[5,23–26] Additionally, interlayer vibronic exciton-phonon states have been reported in a TMD/TMD diode using photocurrent measurements.[27] Our work is distinct in that we report the formation of an interlayer exciton-phonon bound state.[13,14] Exciton-phonon bound states are uncommon quasiparticles that are formed from particles whose number is not conserved (i.e., phonons).[13,14] This means the exciton-phonon bound states are only stable if their real decay is forbidden by the exciton laws of conservation of energy and momentum, during phonon disappearance.[14] While intralayer exciton-phonon bound states have been reported in low-dimensional materials,[12] to our knowledge, such a state has never been observed to form across an interlayer region in a vdW heterostructure.

Both bismuth selenide ($Bi_2Se_3$) and monolayer tungsten disulfide (1L $WS_2$) have independently demonstrated promise for a variety of advanced technologies, including spintronics[18,20] and quantum computing.[21,22] $Bi_2Se_3$ is a topological insulator (TI) with gapped bulk states and gapless surface conducting states that are spin-momentum locked, enabling advanced field effect transistors (FETs),[28] magneto-electric devices,[29] topological qubits,[22] and spin-orbit torque (SOT) devices.[20,30] Monolayer $WS_2$ is a direct band gap semiconductor with independently optically addressable valleys, a strong light-matter interaction, and high sensitivity to surrounding fields, enabling FETs,[31] sensors,[32] valleytronics devices,[21] ferroelectric modulation of exciton populations,[33] and various photonic devices (e.g., LEDs, modulators, lasers).[34] $Bi_2Se_3$/monolayer $WS_2$ heterostructures offer the possibility to combine strong light-matter interaction and spin-locked current into one device.

## Results

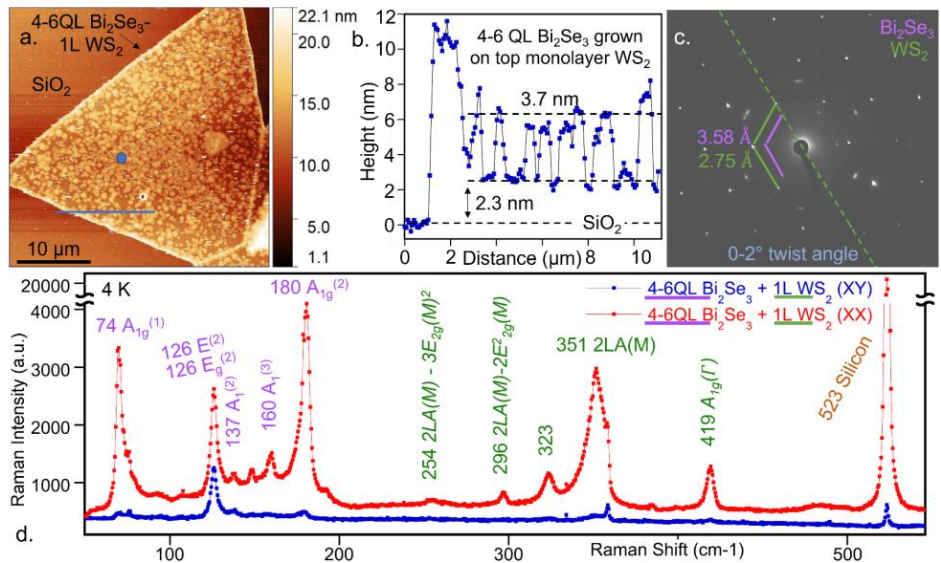

**Figure 1: Characterization of 4-6QL $Bi_2Se_3$ + 1L $WS_2$ 2D heterostructures.** (a) AFM scan and (b) line profile corresponding to blue line in 1a. Blue spot marks a location with continuous film (Section S1) and a representative location studied. (c) TEM SAED image

showing Bi$_2$Se$_3$ grows crystalline on WS$_2$ across a 0-2° twist angle range. (d) Linearly polarized Raman response at 4 K with modes identified. Porto notation is used (see methods).

Figure 1 presents data characterizing the as-grown 4-6 quintuple layer (QL) Bi$_2$Se$_3$ + monolayer (1L) WS$_2$ vdW heterostructure, synthesized using chemical vapor deposition (CVD). Figure 1a-b shows an atomic force microscope (AFM) scan and corresponding line profile, respectively, of a representative sample. For reference, each QL of Bi$_2$Se$_3$ is ~ 1nm thick, and a monolayer of WS$_2$ is 0.7 nm thick. Based on the AFM data and diminished PL intensity, we conclude that Bi$_2$Se$_3$ grew as a uniform QL over 1L WS$_2$, with taller islands (~3.7nm) on top that merge together to form a continuous multilayer film. Section S1 shows additional AFM data and fluorescence measurements showing that the Bi$_2$Se$_3$ is initially continuous, followed by growth of multilayer islands characteristic of Bi$_2$Se$_3$ growth. Transmission electron microscope selected area electron diffraction (TEM-SAED) measurements demonstrate both the WS$_2$ and Bi$_2$Se$_3$ grow crystalline, and that the Bi$_2$Se$_3$ grows within a narrow range of twist angles (0-2°) around 0° aligned with the WS$_2$. Our TEM results are in agreement with previous work,[35] indicating the interlayer interaction is sufficiently strong to induce epitaxial growth. Figure 1d shows linearly polarized Raman measurements taken at 4 K where the well-formed modes correspond closely to the respective materials,[9,36] suggesting good crystalline and stoichiometric sample quality.

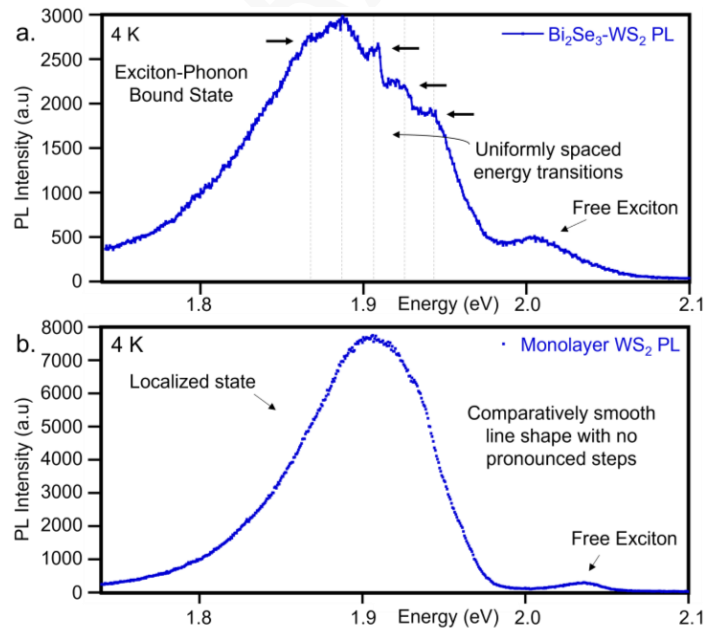

**Figure 2: Manifestation of interlayer exciton-phonon bound state.** (a) 4-6QL Bi$_2$Se$_3$ + 1L WS$_2$ 2D heterostructure PL at 4 K showing the free exciton and localized state. Evenly spaced "steps" are observed, whose energy spacing (19.3meV) corresponds well to the 160cm$^{-1}$ A$_1^{(3)}$ surface phonon. Previous work identified the A$_1^{(3)}$ mode as particularly susceptible to electron-phonon coupling.[9,10] (b) 1L WS$_2$ PL at 4 K, where a comparatively smooth localized state is observed.

Figure 2a shows the PL spectrum of the $Bi_2Se_3$/$WS_2$ heterostructure sample used in Figure 1, which exhibits a red shift compared to the as-grown monolayer $WS_2$ reference sample (Figure 2b). A similar red shift was reported in graphene/$WS_2$ heterostructures, and is attributed to charge transfer.[37] Both spectra are dominated by a broad asymmetric feature approximately 130 meV below the free exciton with a long tail at low energy, generally attributed to contributions from multiple defect-bound exciton complexes.[38–40] Power-dependent (Section S2) and temperature-dependent (Section S3) measurements validate our assignment of exciton species, where the lower energy peak is a localized exciton (or bound exciton), and the higher energy peak is a neutral free exciton.

In addition, the heterostructure PL exhibits regularly spaced peaks not observed in the reference monolayer. The peaks on the high energy side are more prominent, while peaks on the low energy side are likely obscured by the low energy tail of the bound excitons. We use visual inspection to identify these peak positions to determine their energy separation. Efforts to quantitatively extract these positions by analyzing the double derivative extrema yielded comparable results (Section S4). Our analysis found the peaks to be approximately evenly spaced, with an energy spacing of 19.3meV (156cm$^{-1}$), which corresponds to the $Bi_2Se_3$ $A_1^{(3)}$ surface phonon mode (160cm$^{-1}$). Of note, previous work found the $A_1^{(3)}$ mode produces a Fano shape under resonant conditions, indicative of electron-phonon coupling at the $Bi_2Se_3$ surface.[9,10]

Electron-phonon and exciton-phonon bound states (or quasiparticles) have been shown to produce a series of equally spaced features, often overlayed on a larger PL or optical reflection curve, where the period of the emerged features approximately matches the phonon's energy.[11–15,41] As such, the presence of such quasiparticles can be inferred when evenly spaced peaks are observed that correspond to a Raman phonon mode.[13] The number of peaks and consistency of their spacing is material and environment specific, as well as the strength of the exciton-phonon coupling.[12,13] The task is further complicated by the fact that the evenly spaced features are frequently overlayed on a larger peak, which obscures their precise peak position.[11–15] Exciton-phonon bound states are frequently strongest when the exciton is localized (or bound) to an impurity center,[13,14] consistent with our observations where the quasiparticle corresponds to the localized state. We note the formation of an exciton-phonon quasiparticle is distinct from exciton-phonon scattering, in part because the momenta of the exciton and phonon are coupled.[14] As such, exciton-phonon quasiparticle interactions can result in uncommon observations, such as the emission of distinct features above the exciton energy.[12–14]

The energy spacing was independently observed and measured in several samples (see Section S4 for data from additional samples). We therefore attribute these multiple peaks as arising from exciton-phonon coupling to the $Bi_2Se_3$ $A_1^{(3)}$ surface phonon, a mode with an energy of ~19.8 meV that is particularly susceptible to electron-phonon coupling,[9,10] which together form am exciton-phonon quasiparticle. The PL spectrum of as-grown monolayer $WS_2$ shown in Fig 2b does not show such additional peaks, demonstrating that coupling to the $Bi_2Se_3$ is responsible for these dramatic features in the PL.

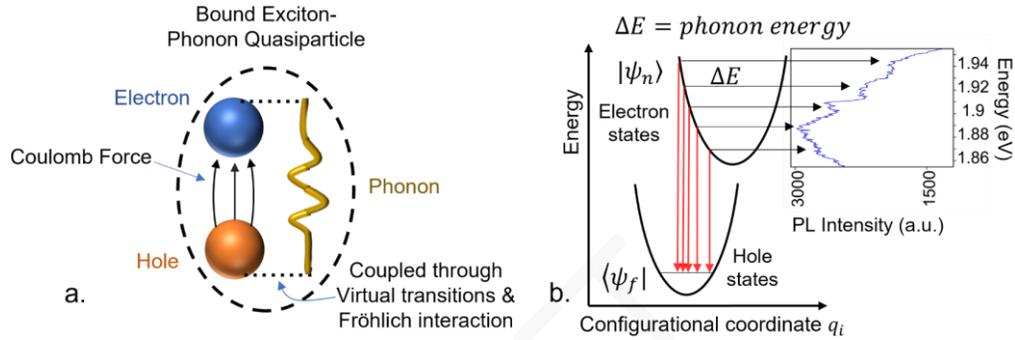

**Figure 3: Schematic of exciton-phonon quasiparticle recombination pathways.** (a) A bound exciton coupled with a phonon to form a bound exciton-phonon quasiparticle, where the phonon interaction modifies the exciton recombination energy. (b) The phonon interaction can induce quantized energy levels corresponding to the phonon mode energy, which leads to multiple recombination pathways that manifest as equally spaced features in experiment. Note, configurational coordinates represent the relative real-space displacement of the initial (electron) and final (hole) states. Inset is data from Figure 2.

Figure 3 presents a graphical explanation of the exciton-phonon bound state quasiparticle,[14] and a mechanism for generating evenly spaced peaks in exciton-phonon systems.[42] The exciton-phonon bound state quasiparticle is only understood by combining several concepts from quantum physics. The quasiparticle is composed of an exciton and a phonon, which are coupled together through virtual transitions that are related to the Fröhlich interaction.[14,42] Virtual transitions are short-lived, unobservable quantum effects that can facilitate a different measurement (or effect), which is physically detectable. For example, Raman spectroscopy is detectable, but often requires virtual states and virtual transitions to manifest, which themselves cannot be observed.[43] The Fröhlich interaction describes the coupling of electrons and phonons through the movement and ionization of a lattice.[44]

We provide two frameworks to qualitatively understand the exciton-phonon mechanism for producing the evenly spaced peaks. First, when the exciton's and phonon's center-of-mass are coupled together, the exciton recoil term and phonon creation-annihilation operators become coupled, thereby modulating exciton recombination energies at quantized phonon intervals.[14] The second framework is shown in Figure 3b, where it assumes the electron and hole are positioned at different locations within the heterostructure, but still bound to form an exciton. As the locations vibrate along a phonon mode, they change atomic coordinates (e.g., configurational coordinates), which alters the overlap of the electron and hole wavefunctions. We assume the Born-Oppenheimer approximation, where the recombination and photon emission is instantaneous compared to the much slower atomic movements. As the electron and hole positions oscillate, the conduction and valence bands change relative coordinates, thereby changing the recombination pathways to different initial and final states, but only at quantized intervals that correspond to the phonon energy.[13,14] Note, configurational coordinate diagrams are plotted in real-space, which displays the valence band as a parabola pointed down.

Previous work found notable interlayer hybridization between $Bi_2Se_3$ and monolayer $WS_2$ that greatly impacts the $WS_2$ excitonic activity,[17,18] strains atoms at the interface,[16] and induces the formation of a pure electronic moiré lattice at the interface.[16] Such interlayer hybridization facilitates the exchange of electrons between the materials and modifies interlayer bonding, thereby setting conditions that encourage the formation of interlayer quasiparticles.[19] We note, the adjacent $Bi_2Se_3$ dramatically reduces the PL intensity of the $WS_2$, due in part to the emergence of non-radiative recombination pathways and charge transfer from the monolayer $WS_2$ into $Bi_2Se_3$.[17,18] We also observe a strong reduction (>2x) in PL intensity. But in addition, our work clearly reveals the presence of these exciton-phonon bound state interactions.

Section S3 show $Bi_2Se_3$-$WS_2$ temperature-dependent measurements, which reveal the localized exciton state (LS) exhibits an unusual negative thermal quenching, where the LS intensity increases and peaks at ~75 K, before decreasing.[45] Previous theory work proposed that as the temperature increases, electrons are thermally excited from nearby bound states into a primary state that allows for radiative recombination. The LS and free exciton peak positions shift lower in energy as the temperature increases, in agreement with the Varshni equation, demonstrating temperature dependent recombination energy.[46] Together, the PL results demonstrate complex behavior, consistent with notable interlayer hybridization between $Bi_2Se_3$ and monolayer $WS_2$.[16,18]

We believe it is unlikely that a $WS_2$ phonon is forming the bound state. First, we observe no indications of an exciton-phonon bound state in 60+ pristine monolayer $WS_2$ samples probed, and we could not find any literature reporting such an effect, suggesting Bi$_2$Se$_3$ grown on top is required to facilitate the quasiparticle. Second, the WS$_2$ phonon modes detected using Raman spectroscopy are likely too high energy. For example, the nearest WS$_2$ Raman mode, LA(M) at ~180cm$^{-1}$ (or 22.3meV), is 15.5% above the measured energy spacing of 19.3meV. Third, WS$_2$ LA(M) is an acoustic mode, which are less likely to form bound states with excitons compared to optical modes.[13] In contrast, the Bi$_2$Se$_3$ A$_1^{(3)}$ surface phonon is only 2.6% above the measured energy spacing, it is an optical mode, and it is particularly susceptible to electron-phonon coupling.[9,10]

It is unlikely that potential defects in WS$_2$ resulting from the Bi$_2$Se$_3$ growth could alone – without the presence of the Bi$_2$Se$_3$ on top – produce the evenly spaced peaks. Limited growth temperature (210°C) and low growth time (27 min.) constrain potential selenium or bismuth doping. Further, selenization (i.e., forming WS$_{2x}$Se$_{2(1-x)}$ alloy) could not be detected in either PL or Raman spectra, suggesting potential selenium doping has a minimal effect on the phonon or exciton modes.[47]

Exciton-phonon coupling and scattering plays a prominent role in monolayer TMDs, where it influences a variety of properties, including valley polarization,[48,49] spectral broadening,[49,50] and mobility.[51] Despite this, phonon scattering alone is unlikely to produce periodic features (or vibronic transitions), and a more complex mechanism is required. For example, the Franck-Condon Principle (FCP) enables multiple quantized radiative recombination pathways.[52] However, it is primarily applied to small molecules, whose excited states shift the atomic coordinates.[53] Conversely, the exciton-phonon quasiparticle is a theoretical framework designed for crystals,[13,14] where phonon transitions are central, leading us to conclude an interlayer exciton-phonon quasiparticle is the most appropriate assignment.

It should be noted that the FCP can be applied to crystalline solids when excitons self-localize and a strong electron-phonon coupling is present.[54,55] Evidence is emerging that conditions in TMD-based van der Waals heterostructures may be conducive to self-localized excitons,[56,57] raising the prospect that the free-excitons in WS$_2$ are self-trapping when forming the exciton-phonon bound state. Further research is required to determine the model that best describes the system.

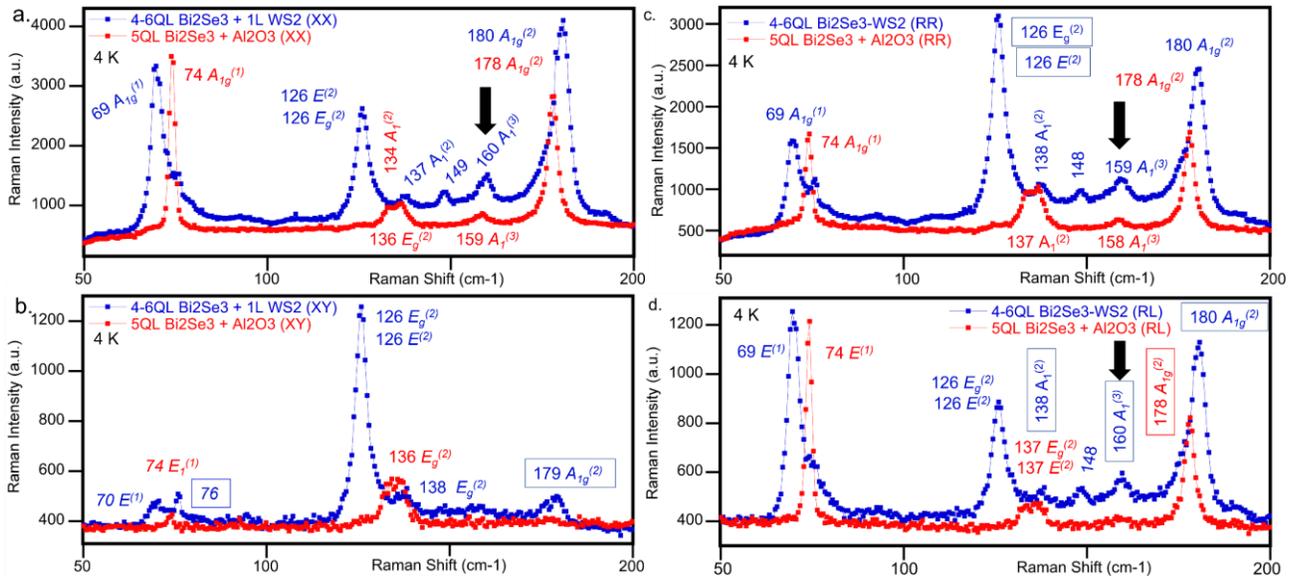

**Figure 4: Prominent surface phonons and local symmetry breaking.** (a)-(b) Linear and (c)-(d) circular polarized Raman response from CVD-grown $Bi_2Se_3$-$WS_2$ heterostructure and MBE-grown $Bi_2Se_3$-$Al_2O_3$ (reference sample). Black arrow identifies the $A_1^{(3)}$ surface phonon, which is associated with the interlayer exciton-phonon bound state. Phonon modes forbidden in a specific scattering geometry (i.e., XX, XY, RR, or RL) are labeled with a boxed-outline.[9] The presence of $A_1^{(2)}$ and $A_1^{(3)}$ surface phonons in a forbidden scattering geometry (i.e., RL), suggests the local symmetry breaking at the surface,[9] consistent with a strong $WS_2$ interlayer coupling. Porto notation is used (see methods).

To elucidate the interlayer exciton-phonon coupling, Figure 4 shows linear and circular polarization Raman measurements at 4 K for CVD-grown $Bi_2Se_3$-$WS_2$, and a $Bi_2Se_3$ reference sample, grown using molecular beam epitaxy (MBE) on $Al_2O_3$. Previous work demonstrated the high quality of the MBE-grown $Bi_2Se_3$ in our setup.[58] Polarized Raman spectroscopy is a reliable method to identify and distinguish different phonon modes with greater confidence because the Raman response can be separated from each symmetry channel.[9,59,60] More specifically, each crystallographic point group contains symmetries that can be used to calculate the reduced Raman tensors, which describe scattering of phonon modes.[60] Applying polarization tensors to Raman tensors reveals the scattering efficiency of phonon modes exposed to polarized light.

Figure 4a-b (Figure 4c-d) show the linearly (circularly) polarized Raman measurements. Section S5 contains expanded information on the Raman measurements, including each phonon mode's attributes and a literature comparison. Crystal symmetry forbids certain Raman scattering in $Bi_2Se_3$.[60] If the symmetry is disrupted, these modes can become allowed.[9,60] The presence of $A_1^{(2)}$ and $A_1^{(3)}$ surface phonons in a forbidden scattering geometry (i.e., RL), suggests the local symmetry breaking at the surface,[9] consistent

with a strong WS$_2$ interlayer coupling. Additionally, the A$_1^{(2)}$ and A$_1^{(3)}$ surface phonon modes are markedly more intense, consistent with an increased phonon population.

Compared to the 5QL reference sample, as well as bulk Bi$_2$Se$_3$ from literature,[9] we observe forbidden modes at greater intensity in Bi$_2$Se$_3$-WS$_2$ heterostructures, suggesting the crystal symmetry is being disrupted to a greater degree. For example, the A$_{1g}^{(2)}$ mode is forbidden in the XY configuration. While the reference sample shows no detectable peak, we observe it prominently in the Bi$_2$Se$_3$-WS$_2$ heterostructure. Of note, although all the peaks observed correspond well to either Bi$_2$Se$_3$ or monolayer WS$_2$ Raman modes,[9,36] we cannot exclude the possibility that combination[61] or moiré Raman modes[62] emerge at overlapping wavenumbers.

Numerous effects influence local crystalline symmetry, including defects, strain, and substrate effects. Previous work found evidence of strain at the Bi$_2$Se$_3$-WS$_2$ interface and a purely electronic 2D lattice between the materials that alters the surrounding electronic environment.[16] Such effects impact the breakdown of translational symmetry at the interface, which subsequently affects surface phonon modes.[63]

Our Raman measurements are consistent with a strong interlayer coupling that encourages strain and charge redistribution at the interface. Together, notable interlayer hybridization[16,18] and a strong interlayer coupling form an interfacial environment that facilitates interlayer interactions of different particles, thereby encouraging the formation of exciton-phonon quasiparticles.[19] Additionally, our measurements demonstrate a rich landscape of phonons, including comparatively intense surface phonons, which are consistent with an increased phonon population.

**Conclusions**

We have demonstrated formation of an interlayer exciton-phonon quasiparticle (or exciton-phonon bound state) between localized excitons in monolayer WS$_2$ and the Bi$_2$Se$_3$ A$_1^{(3)}$ surface phonon, a mode particularly susceptible to electron-phonon coupling.[9,10] We detect evenly spaced features in the PL spectrum with an energy separation that matches the Bi$_2$Se$_3$ A$_1^{(3)}$ surface phonon, overlayed on the WS$_2$ localized exciton emission peak. Polarized Raman spectroscopy detects forbidden Bi$_2$Se$_3$ surface phonon modes, suggesting broken crystalline symmetry at the surface. While several publications have reported exciton/electron-phonon coupling across the interlayer region,[5,23–26] our work is distinct in that we report the formation of an interlayer exciton-phonon bound state, an uncommon quasiparticle composed of phonons, whose particle number is not conserved.[13,14] $Bi_2Se_3$/monolayer $WS_2$ heterostructures offer the possibility to combine strong light-matter interaction and spin-locked current into one material. Understanding the interlayer coupling is central to elucidating how their combined properties evolve, enabling devices for spintronics,[18,20] valleytronics,[21] and quantum computing[21,22] applications.

**Methods**

**Material Growth – Few-layer $Bi_2Se_3$:** The $Bi_2Se_3$ films were grown on 10 × 10 mm$^2$ c-plane (0001) sapphire ($Al_2O_3$) substrates using molecular beam epitaxy (MBE) with base pressure below 5 × 10$^{-10}$ Torr. The substrates were initially annealed ex-situ at 1,000 °C under the atmospheric pressure, and ozone cleaned in-situ under 200 Torr of oxygen pressure. It is then annealed at 600 °C for 20 min in the ultra-high vacuum MBE chamber. Individual sources of high-purity (99.999%) Bi and Se were evaporated from standard effusion cells during the film growth. Se flux was maintained at least ten times higher than Bi's to minimize Se vacancies. To obtain an atomically sharp interface between the $Bi_2Se_3$ layer and the substrate, we adopted the two-step growth scheme.[64] First, the initial 3 QL $Bi_2Se_3$ is grown at 170 °C. It is slowly annealed to 300 °C, and followed by deposition of the remaining 2QL $Bi_2Se_3$ layers. 5QL $Bi_2Se_3$ was grown.

**Material Growth – $Bi_2Se_3$-$WS_2$ 2D Heterostructure:** Monolayer $WS_2$ is synthesized at ambient pressure in 2-inch diameter quartz tube furnaces on $SiO_2$/Si substrates (275 nm thickness of $SiO_2$). Prior to use, all $SiO_2$/Si substrates are cleaned in acetone, IPA, and Piranha etch ($H_2SO_4$+$H_2O_2$) then thoroughly rinsed in DI water. At the center of the furnace is positioned a quartz boat containing ~1g of $WO_3$ powder. Two $SiO_2$/Si wafers are positioned face-down, directly above the oxide precursor. A separate quartz boat containing sulfur powder is placed upstream, outside the furnace-heating zone. The upstream $SiO_2$/Si wafer contains perylene-3,4,9,10-tetracarboxylic acid tetrapotassium salt (PTAS) seeding molecules, while the downstream substrate is untreated. The hexagonal PTAS molecules are carried downstream to the untreated substrate and promote lateral growth of monolayer $WS_2$. Pure argon (65 sccm) is used as the furnace heats to the target temperature. Upon reaching the target temperature in the range of 825 to 875 °C, 10 sccm $H_2$ is added to the Ar flow and maintained throughout the 10-minute soak and subsequent cooling to room temperature.

4-6QL $Bi_2Se_3$ was grown on top of monolayer $WS_2$ using chemical vapor deposition (CVD) in a two-zone furnace with a 2″ quartz tube. High-purity $Bi_2Se_3$ flakes are ground using a mortar and pestle into a fine dust. The powdered $Bi_2Se_3$ is placed in a ceramic boat and inserted into the furnace's quartz tube, and pushed into the center of the furnace's first zone. The monolayer $WS_2$, which is on an $SiO_2$ substrate, is placed downstream of the $Bi_2Se_3$ into the center of the furnace's second zone. The furnace is pumped down to ~20mTorr. An argon (Ar) carrier gas is flown into the furnace at 80sccm. The $Bi_2Se_3$ is heated to 520°C, and the $WS_2$ is heated to 210°C. The ramp rate is ~55°C/min, and the total growth is 27 min.

**Raman and photoluminescence measurements at low temperature:** A Horiba LabRARM HR Evolution with both linear and circular polarization attachments, and a low-temperature Montana cryostat, was used for Raman and photoluminescence (PL) spectroscopy measurements. We use Porto Notation (i.e., $\bar{z}(\mu\nu)z$) where μ(ν) is the incident (scattered) polarization, and $\bar{z}(z)$ is the incoming (outgoing) direction. We define $R = X + iY$ and $L = X - iY$. Previous work showed that laser exposure of monolayer materials at low temperature can anneal and laser-dope them.[65,66] We attempted to mitigate this using very low powers (~320nW) and short exposure times (~30s) for most measurements. We verified that the laser exposure had a minimal effect on the material by collecting multiple successive spectra. A long-distance 50x objective was used with a laser spot diameter of ~1.9 μm at the lowest powers.

**Transmission Electron Microscopy:** $Bi_2Se_3$-$WS_2$ 2D heterostructures were transferred onto a holey amorphous $SiN_x$ TEM grid using the water-assisted-pick-up transfer method.[67] Selected area electron diffraction (SAED) were performed with a JEOL JEM2200FS operating at 200 kV, equipped with a high-speed Gatan OneView camera. The SAED patterns were internally calibrated to the $WS_2$, and an aperture of approximately 200nm was used. We suspect that the wet transfer of $Bi_2Se_3$-$WS_2$ 2D heterostructures partially disrupts the crystal order, possibly due to a combination of the force and liquids applied during the transfer process.

**Photoluminescence Spectra computational analysis and fitting:** All code was written in Python using the Spyder integrated development environment (IDE). Spyder belongs to the MIT License and is distributed through the Anaconda environment. The curve_fit() function with a variety of initial values and boundary conditions were used to verify fit robustness. The fitting was corroborated using cross-validation, where we uniformly removed 20% of the data points from a spectrum and repeated fitting. No notable changes to the fitting were detected, suggesting noise is not skewing the fit.
The energy location of each feature was extracted by taking the double derivative of the best fit, and finding the minimum values. The minimum values of the double derivative matched well with the peak locations of Lorentzian functions, reinforcing our method to quantitatively extract the features.

**Acknowledgments**
We thank Dr. Darshana Wickramaratne at the Naval Research Laboratory for their insight and fruitful discussions.

**Supporting Information**
Additional characterization of representative regions probed (Section S1); Power-Dependent Measurements (Section S2); Temperature-Dependent Measurements (Section S3); Expanded Analysis of the Interlayer Exciton-Phonon Bound State (Section S4); Expanded analysis of low-temperature (4 K) $Bi_2Se_3$ phonon modes (Section S5);

# Supporting Information

**Interlayer Exciton–Phonon Bound State in Bi$_2$Se$_3$/monolayer WS$_2$ van der Waals Heterostructures**

Zachariah Hennighausen[1,*], Jisoo Moon[1], Kathleen M. McCreary[2], Connie H. Li,[2] Olaf M.J. van 't Erve[2], and Berend T. Jonker[2,*]

[1] NRC Postdoc at the Materials Science and Technology Division, Naval Research Laboratory, Washington, D.C. 20375, USA
[2] Materials Science and Technology Division, Naval Research Laboratory, Washington, D.C. 20375, USA

**Section S1. Additional characterization of representative regions probed**

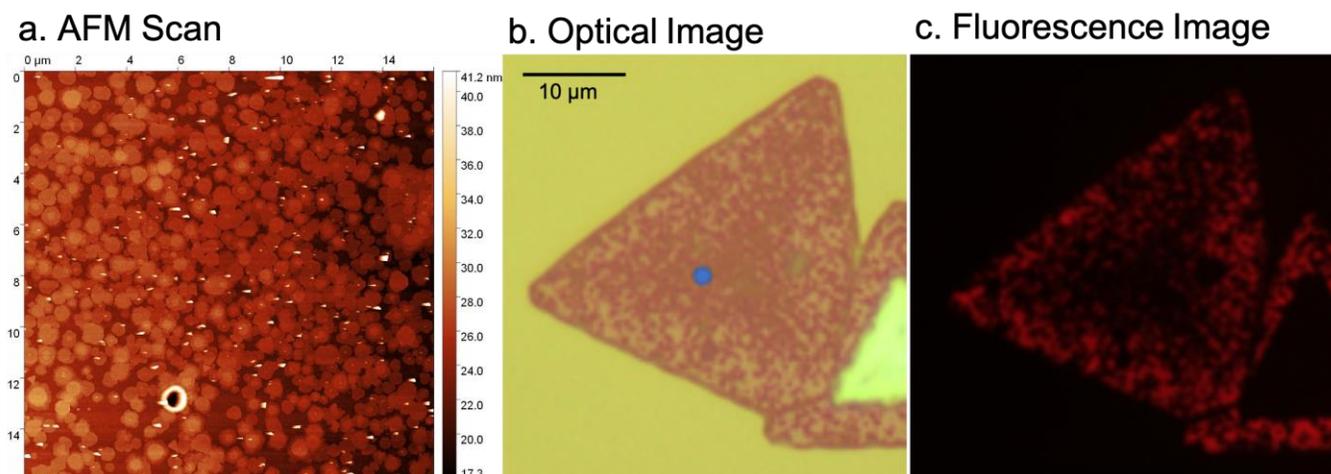

**Figure S1. Bi$_2$Se$_3$ islands grow together forming a nearly continuous 4-6QL film on monolayer WS$_2$.** (a) AFM scan around location of blue spot in (b). (b) optical image where blue spot identifies representative location probed. (c) Fluorescence image showing that the photoluminescence (PL) is fully quenched at the center, indicating the Bi$_2$Se$_3$ is a nearly 4-6QL continuous film.

Previous work found that forming heterostructures by growing 1QL Bi$_2$Se$_3$ on a monolayer (1L) transition metal dichalcogenide (TMD) significantly reduces the TMD PL intensity compared to their as-grown (i.e., bare) 1L TMDs counterparts,[1–4] likely due in part to the formation of an indirect bandgap and static charge transfer. Further, it was found that as additional Bi$_2$Se$_3$ layers are grown on top, the PL continues to diminish and quench because the bandgap becomes increasingly indirect with increasing Bi$_2$Se$_3$.[4] This effect is comparable to the PL intensity evolution as a TMD layer count increases (i.e., monolayer vs. bilayer vs. trilayer). More specifically, increasing layer count from 1L to 2L dramatically reduces the PL intensity, and increasing from 2L to 3L further diminishes the PL intensity. As layer count increases, the sample approaches the bulk properties.

## Section S2. Power-dependent measurements

Section S2 shows power-law fitting to power-dependent measurements, which is used to identify the species of exciton each peak originates from (e.g., localized state, free exciton, biexciton). The localized state (LS) has a coefficient of 0.688, while the free exciton (FE) is 0.975, and the ratio of localized-to-free exciton intensity decreases with increasing power, enabling us to label the excitons with high confidence. Note, there are a diverse number of localized and bound excitons, whose classification depends in part on local chemistry and the spatial extent of the wavefunction. We cannot identify definitively the type of localized exciton.

At low powers, the ratio of localized to free exciton is greater. However, as the power increases, the ratio decreases because the number of electrons excited from the valence band to the conduction band increases. As the electrons recombine, they begin to saturate the localized states, pushing a higher ratio of electrons in the free exciton states (Figure S2), which are less easily saturated.[5,6]

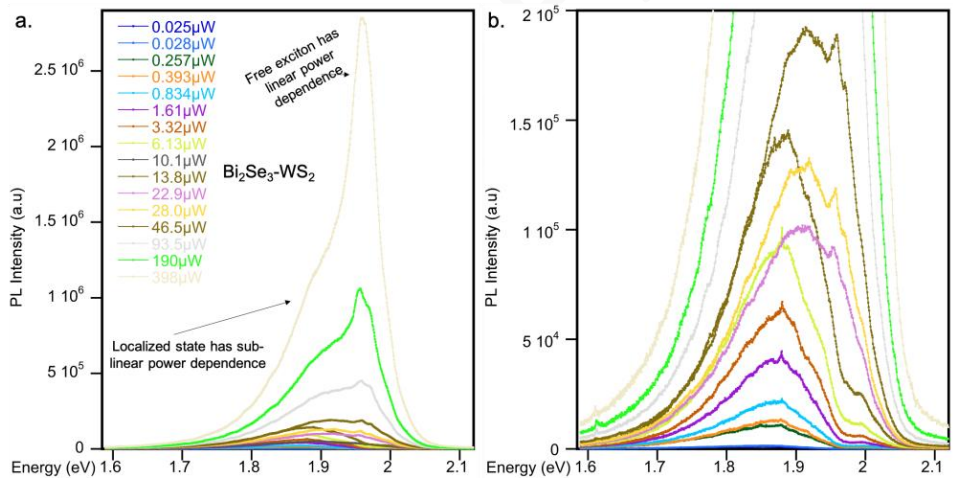

**Figure S2.** 4-6QL $Bi_2Se_3$ + monolayer $WS_2$ 2D heterostructure power-dependent measurements.

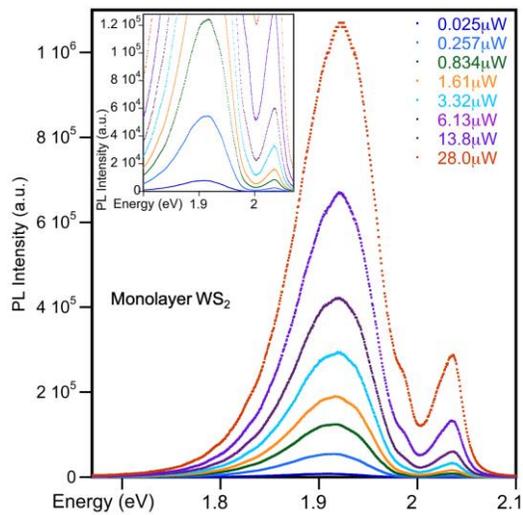

**Figure S3. As-grown monolayer WS₂ power-dependent measurements.** Graphed evolution or curve peaks is shown below in Figure S4.

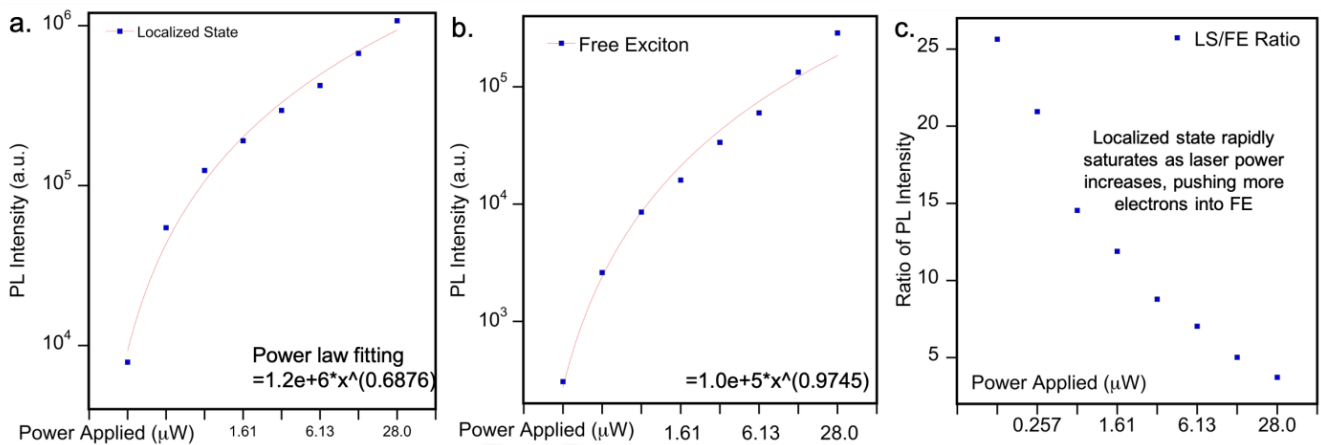

**Figure S4. As-grown monolayer WS₂ power-dependent measurements: Peak intensities plotted from Figure S3.** The expected behavior validates the assignment of the localized and free excitons. (a) Shows the localized state exciton (LS) and (b) the free exciton (LE), which have a fitting coefficient of less than one and approximately equal to one, respectively. The ratio of LS/FE decreases with laser power. Together, the results provide high confidence of the locations of the localized and free exciton states.[5,6]

The peak intensities were extracted using Spyder integrated development environment (IDE) and the curve_fit() function. The results were corroborated with manual inspection of the highest intensity pixel recorded by the equipment for each exciton curve.

# Section S3. Temperature-dependent measurements

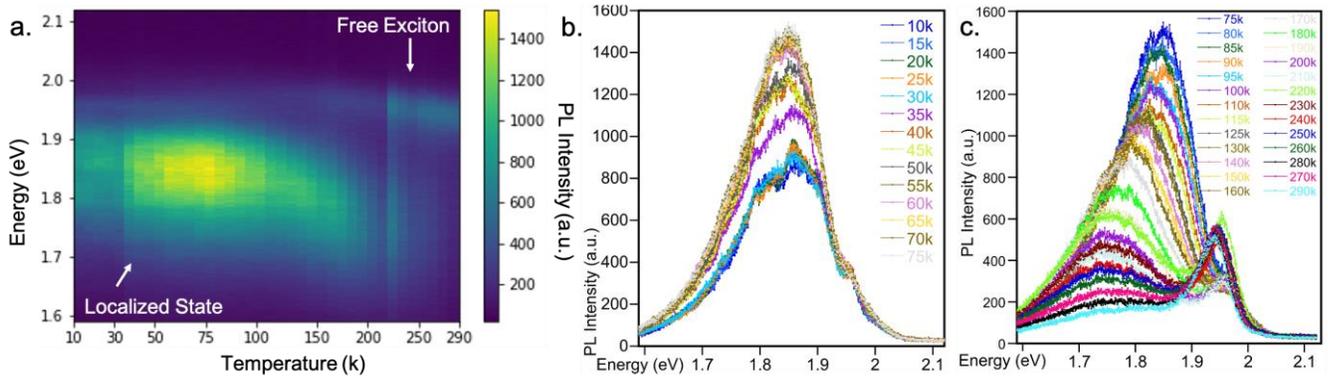

**Figure S5. 4-6QL Bi$_2$Se$_3$ + monolayer WS$_2$ 2D heterostructure temperature-dependent measurements.** (a) 2D density plot of the PL spectra with temperature, showing the localized and free exciton states. (b)-(c) PL spectra as a function of temperature. Interestingly, negative thermal quenching is observed for the localized state, an unusual phenomena where the PL intensity increases with temperature.[7] Previous theory work proposed that as the temperature increases, electrons are thermally excited from nearby bound states into a primary state that allows for radiative recombination.[7] The peak position shifts lower as the temperature increases, in agreement with the Varshni equation, formalism used to describe the PL evolution in semiconductors.[8]

# Section S4. Expanded analysis of the interlayer exciton-phonon bound state

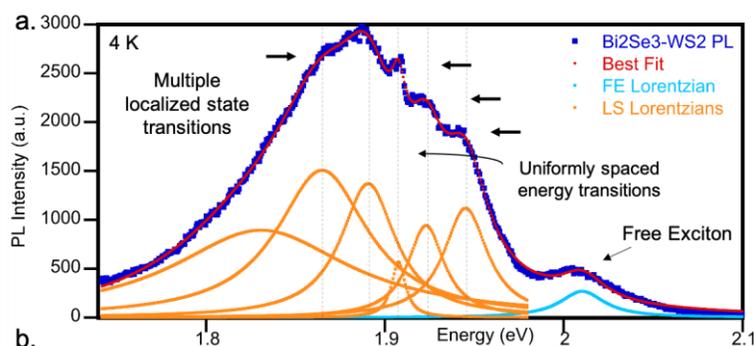

**Figure S6. Bi₂Se₃-WS₂ data from Figure 2 fitted with Lorentzian functions.** We obtain a function that follows the curve and captures the data form. Inflection points were quantitatively identified by taking a double derivative of the best fit. Note, we only use the fitting methodology as a tool to quantitatively extract the inflection points. We do not extract insight from the Lorentzian fitting.

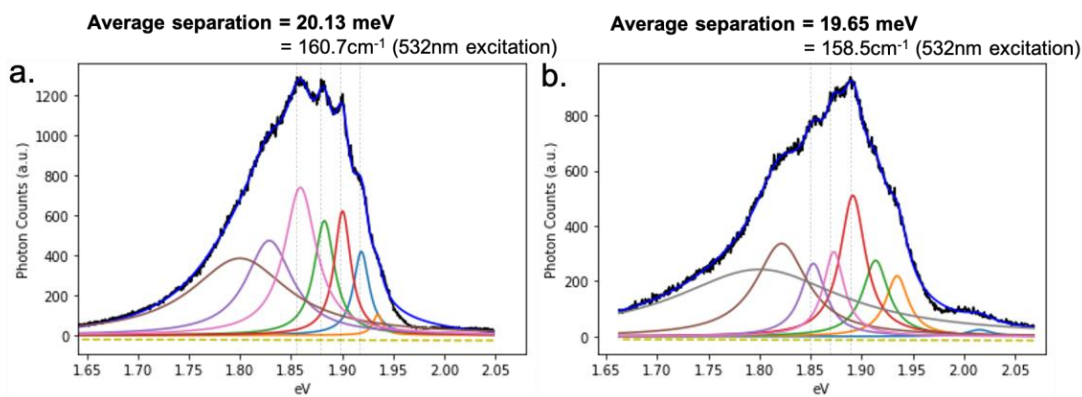

**Figure S7. Representative data with peaks that are approximately equally spaced.**

**Table 1. Spacings between Double Derivative Extrema.** Multiple Lorentzian functions were fit to the data using Python Software (see methods). The double derivative was taken of the best fit function and the minimum extrema were identified. The spacing between the extrema is calculated.

| Spacing | Figure S6 | Figure S7a | Figure S7b |
|---------|-----------|------------|------------|
| 1st     | 20.75     | 18.49      | 18.87      |
| 2nd     | 16.79     | 18.06      | 20.5       |
| 3rd     | 18.13     | 23.85      |            |
| 4th     | 21.54     |            |            |
|         |           |            |            |
| Average | 19.3meV   | 20.1meV    | 19.7meV    |

Methodology for Measuring Spacing of the Interlayer Exciton-Phonon Bound State Features:

The exciton-phonon bound state (or exciton-phonon quasiparticle) presents itself as evenly spaced peaks or features, where the spacing is approximately the phonon energy. As such, the presence of an exciton-phonon or electron-phonon quasiparticle can be inferred when evenly spaced peaks are observed that correspond to a Raman phonon mode.[9–14] The number of peaks and consistency of the even spacing is material and environment specific, as well as the strength of the exciton-phonon coupling.[10,11] Further, the task is complicated by the fact that the evenly spaced features are frequently overlayed on a larger peak, which obscures their precise peak position.[9–14]

When analyzing the PL spectra for a possible exciton-phonon bound state, we start by identifying possible features that could be peaks and then measuring their spacing. To the best of our knowledge, the community primarily relies on visual inspection when identifying and measuring possible exciton-phonon peaks. In this work, we applied two methods for peak identification. The primary method is visual inspection, while the secondary method is a quantitative analysis of best fit double derivative extrema. Note, we only use the fitting methodology as a tool to quantitatively extract the inflection points. We do not extract insight from the Lorentzian fitting.

Only if both methods yielded the same result, did we label the feature as a peak and measure spacings. We fit the experimental data with either 7 or 8 Lorentzian functions, which was sufficient to obtain a best fit that corresponded well to the data moving average. We then took the double derivative of the best fit and identified the minima extrema. We verified each double derivative minima corresponded to a clear visual feature. We then measured the spacing between each double derivative minima. The double derivative extrema frequently corresponded to the location of Lorentzian peaks, but did not overlap exactly.

# Section S5. Expanded analysis of low-temperature (4 K) $Bi_2Se_3$ phonon modes

Table S2. Summary of low-temperature $Bi_2Se_3$ phonon modes for $Bi_2Se_3$-$WS_2$ and $Bi_2Se_3$-$Al_2O_3$. Increased symmetry breaking was observed in $Bi_2Se_3$-$WS_2$ heterostructures compared to $Bi_2Se_3$-$Al_2O_3$.

| Phonon Mode | Bulk/Surface | Scattering Geometry | $Bi_2Se_3$-$WS_2$ (cm-1) | $Bi_2Se_3$-$Al_2O_3$ (cm-1) | Literature: Bulk $Bi_2Se_3$-$Al_2O_3$ (cm-1) | Notable Symmetry breaking |
|---|---|---|---|---|---|---|
| $A_1^{(2)}$ | Surface | RR & XX | 137 | 137 | 136[15] & 129[16] | Yes (RL Channel) |
| $A_1^{(3)}$ | Surface | RR & XX | 160 | 159 | 158[15] & 160[16] | Yes (RL Channel) |
| $A_{1g}^{(1)}$ | Bulk | RR & XX | 69 | 74 | 75[15] & 73[16,17] & 72[18] | No |
| $A_{1g}^{(2)}$ | Bulk | RR & XX | 180 | 178 | 180[15] & 175[16,17] & 174[18] | Yes (XY Channel) |
| $E^{(1)}$ | Surface | RL | 69 | 74 | 67[15] & 68[16] | No |
| $E^{(2)}$ | Surface | RL | 126 | 137 | 126[15] & 125[16] | Yes (RR Channel) |
| $E_g^{(2)}$ | Bulk | RL | 126 | 136 | 137[15] & 133[16] & 131[17,18] | Yes (RR Channel) |

The above Raman modes are identified with the assistance of previous work by matching them to the wavenumber and polarization response, which reveals the symmetry channel.[15,19,20] We primarily relied upon *Kung et al.* and *Gnezdilov et al.* for understanding the symmetry channels and polarized Raman spectroscopy results.[15,16] Our data presented HERSE for low-dimensional $Bi_2Se_3$ using polarized Raman spectroscopy and at low-temperatures are among the few in the literature. While our samples were between 4-6QL $Bi_2Se_3$, the literature values are from bulk $Bi_2Se_3$ samples.

$Bi_2Se_3$ has the $D_{3d}$ point group symmetry in the rhombohedral crystal structure, suggesting one or more of the following point symmetry groups is being broken: $S_6$, $D_3$, $C_{3v}$, $C_3$, $C_{2h}$, $C_2$, $C_s$, or $C_i$.[21] Definitively identifying which symmetry point groups are being disrupted is beyond the scope of this work.

s